\documentclass[aps,prb,showpacs,twocolumn,citeautoscript,superscriptaddress]{revtex4}
\usepackage{amsmath}
\usepackage{graphicx}
\usepackage{dcolumn}
\usepackage{float}

\begin{document}
\title{Reentrant Superconductivity in Eu(Fe$_{1-x}$Ir$_{x}$)$_{2}$As$_{2}$}
\author{U. B. Paramanik}
\author{Debarchan Das}
\author{R. Prasad}
\author{Z. Hossain}
\email{zakir@iitk.ac.in}
\affiliation{Department of Physics, Indian Institute of Technology, Kanpur 208016, India}

\date{\today}

\begin{abstract}
The interplay between superconductivity and Eu$^{2+}$ magnetic ordering in Eu(Fe$_{1-x}$Ir$_{x}$)$_{2}$As$_{2}$ is studied by means of electrical transport and magnetic measurements. For the near optimally doped sample Eu(Fe$_{0.86}$Ir$_{0.14}$)$_{2}$As$_{2}$, we witnessed two distinct transitions : a superconducting transition below 22.6 K which is followed by a resistivity reentrance caused by the ordering of the Eu$^{2+}$ moments. Further, the low field magnetization measurements show a prominent diamagnetic signal due to superconductivity which is remarkable in presence of a large moment magnetically ordered system. The electronic structure for a 12.5\% Ir doped EuFe$_{1.75}$Ir$_{0.25}$As$_{2}$ is investigated along with the parent compound EuFe$_{2}$As$_{2}$. As compared to EuFe$_{2}$As$_{2}$, the doped compound has effectively lower value of density of states throughout the energy scale with more extended bandwidth and stronger hybridization involving Ir. Shifting of Fermi energy and change in band filling in EuFe$_{1.75}$Ir$_{0.25}$As$_{2}$ with respect to the pure compound indicate electron doping in the system.

\end{abstract}

\pacs {74.70.Xa, 74.25.F-, 74.25.Ha, 74.25.Jb}

\maketitle

The interplay between superconductivity (SC) and long range magnetic order has been an interesting topic in condensed matter physics due to various interesting phenomena arising from it including reentrance of superconductivity. Earlier, a few theoretical studies pointed out that a long range ferromagnetic (FM) order or FM impurity inside a material would drastically suppress the superconductivity, while SC and antiferromagnetism (AFM) may coexist.\cite{Ginzburg, Abrikosov, Baltensperger} Decades later the competitive nature between SC and ferromagnetism were first demonstrated in ErRh$_{4}$B$_{4}$ (Ref. 4) and in Ho$_{1.2}$Mo$_{6}$S$_{8}$ (Ref. 5) where SC is destroyed at the onset of long range magnetic order of rare-earth ions. The interplay between SC and long range magnetic order were also investigated in quaternary rare-earth polycrystalline borocarbide system where SC may coexist or compete with long range antiferromagnetic (AFM) order or other kinds of magnetic order with a FM component, e.g in RNi$_{2}$B$_{2}$C (R = Dy, Ho, Er, and Tm).\cite{Eisaki, Goldman, Sinha} The reentrant feature is observed only in presence of external magnetic
field in single crystal of RNi$_{2}$B$_{2}$C (Ref. 9, 10).  In contrast, superconductivity was observed on the border of ferromagnetism in heavy fermion system UGe$_{2}$ (Ref. 11) where SC and band ferromagnetism arise from the same electrons. This makes the interplay between SC and ferromagnetism even more interesting.

The discovery of unconventional superconductivity in proximity to magnetism in iron pnictides once again has given us the opportunity to investigate the issue of the interplay between superconductivity and magnetism. In these itinerant electron systems, the Fe spin density wave (SDW) order and superconductivity compete each other and sometimes coexist. \cite{Wiesenmayer, Drew} We are interested in a system where alongside the SDW ordering in the Fe site, a large localized magnetic moment also subsists in the same system as in case of RNi$_{2}$B$_{2}$C. (Ref. 14) We find that EuFe$_{2}$As$_{2}$ (Ref. 15) is the most convenient system where Eu$^{2+}$ orders antiferromagnetically at 19 K. The interplay between SC and Eu$^{2+}$ magnetism has already been studied in Eu$_{1-x}$K$_{x}$Fe$_{2}$As$_{2}$ (Ref. 16, 17), EuFe$_{2}$(As$_{1-x}$P$_{x}$)$_{2}$ (Ref. 18) and in EuFe$_{2}$As$_{2}$ under applied pressure \cite{Hossain, Matsubayashi, TERASHIMA} where Eu$^{2+}$ moments order antiferromagnetically which can coexists with superconductivity. But there are limited examples of doping in the Fe site where superconductivity coexists with the Eu$^{2+}$ magnetic order. EuFe$_{2-x}$Co$_{x}$As$_{2}$ is one of the rare examples where Eu$^{2+}$ has been proposed to be ordered helically which can coexist with superconductivity.\cite{Jiang, Hu} Although, in this system, superconductivity is only manifested in resistivity but no diamagnetic signal has been observed because of the proximity of Eu$^{2+}$ magnetic ordering. On the other hand, Ni doping in EuFe$_{2-x}Ni_{x}As_{2}$ (Ref. 24) showed only FM ordering of the Eu$^{2+}$ moments but no superconductivity.  However, $4d$ and $5d$ transition metals are quite different from their $3d$ counterparts in several aspects. Since 4d and 5d orbitals are more extended than the $3d$ orbitals, there will be more hybridization with As as well as with Fe. So, the effective Hund's coupling on the atoms will be weaker, which works against magnetism and thereby suppresses the SDW in favor of superconductivity.\cite{Lijun} There are reports on Ir doped ``122" systems which give highest superconducting transition temperature ($T_c$) among the transition metal doped ``122" systems.\cite{Wang, Han} This has motivated us to dope Fe by Ir in EuFe$_{2}$As$_{2}$ which might exhibit $T_{c}$ higher than the Eu$^{2+}$ magnetic ordering temperature.

In this paper, we study the interplay between SC and magnetism in Ir doped  EuFe$_{2}$As$_{2}$ polycrystalline samples through electrical resistivity $\rho(T)$ and magnetization measurements. We observe a sharp resistivity drop below 22.6 K which is ascribed to a SC transition. On further reducing the temperature, $\rho(T)$ increases again and exhibits a maximum at 15 K caused by the ordering of the Eu$^{2+}$ moments. Interestingly, we notice a prominent diamagnetic signal in the low field magnetization measurements.

The polycrystalline samples of Eu(Fe$_{1-x}$Ir$_{x}$)$_{2}$As$_{2}$ ($x$ = 0, 0.05, 0.11 and 0.14) were prepared using solid state reaction method as described in our earlier reports. \cite{Anupam, Anupam1, Hossain1} Stoichiometric amounts of the starting elements of Eu chips (99.9\%), Fe powder (99.999\%), Ir powder (99.99\%) and As chips (99.999\%) were used for the reaction. The crushed polycrystalline samples were characterized by x-ray diffraction with Cu-$K_\alpha$ radiation to determine the single phase nature and crystal structure. Scanning electron microscope (SEM) equipped with energy dispersive x-ray (EDX) analysis was used to check the homogeneity and composition of the samples. The electrical transport properties were measured by standard four probe technique using Physical Properties Measurement System (PPMS, Quantum Design, USA) and close cycle refrigerator (Oxford Instruments). The magnetic properties of the samples have been probed using a commercial SQUID magnetometer (MPMS, Quantum-Design).

\begin{figure}[htb!]
\includegraphics[width=8cm, keepaspectratio]{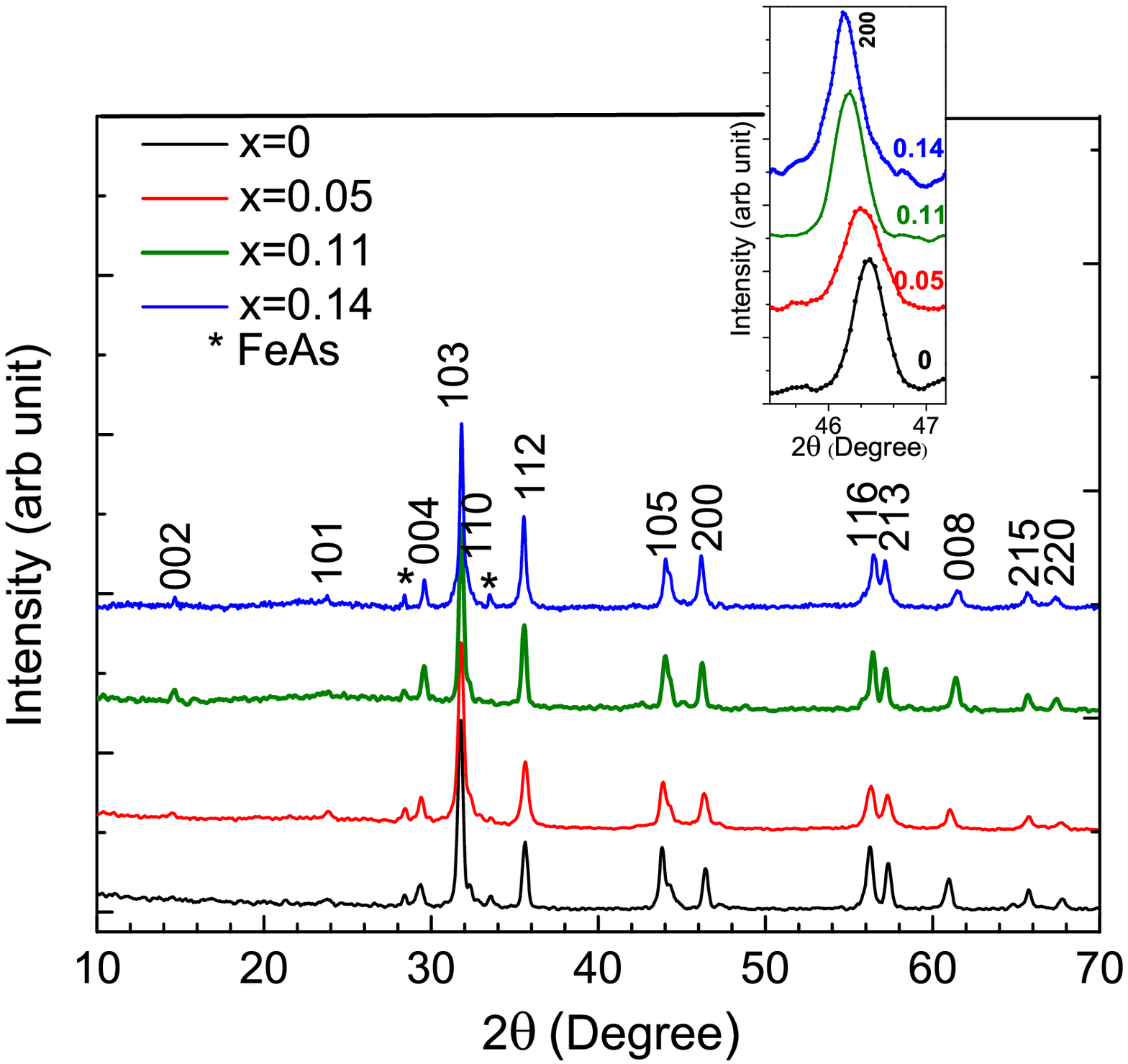}
\includegraphics[width=4.3cm, keepaspectratio]{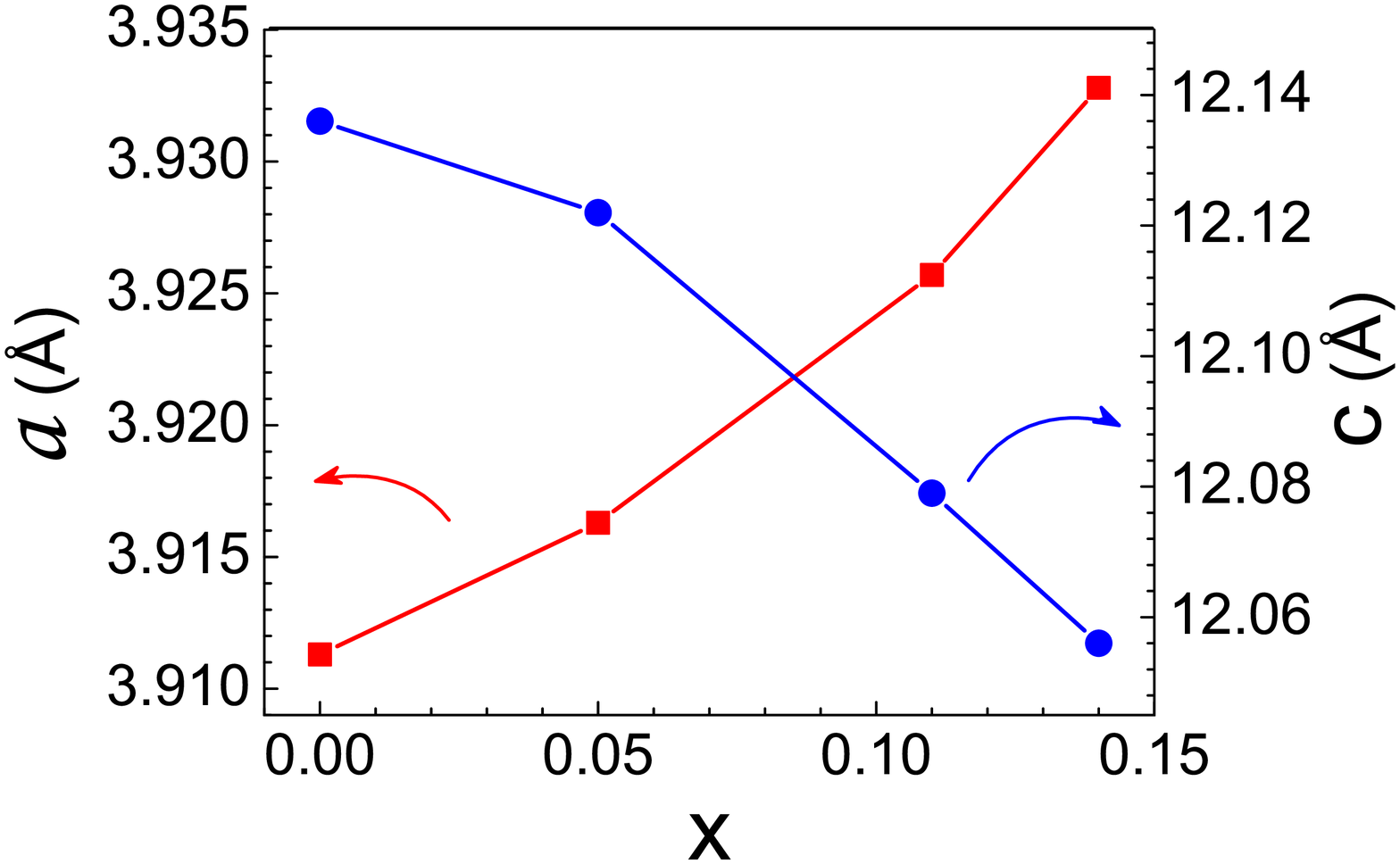}
\includegraphics[width=4.2cm, keepaspectratio]{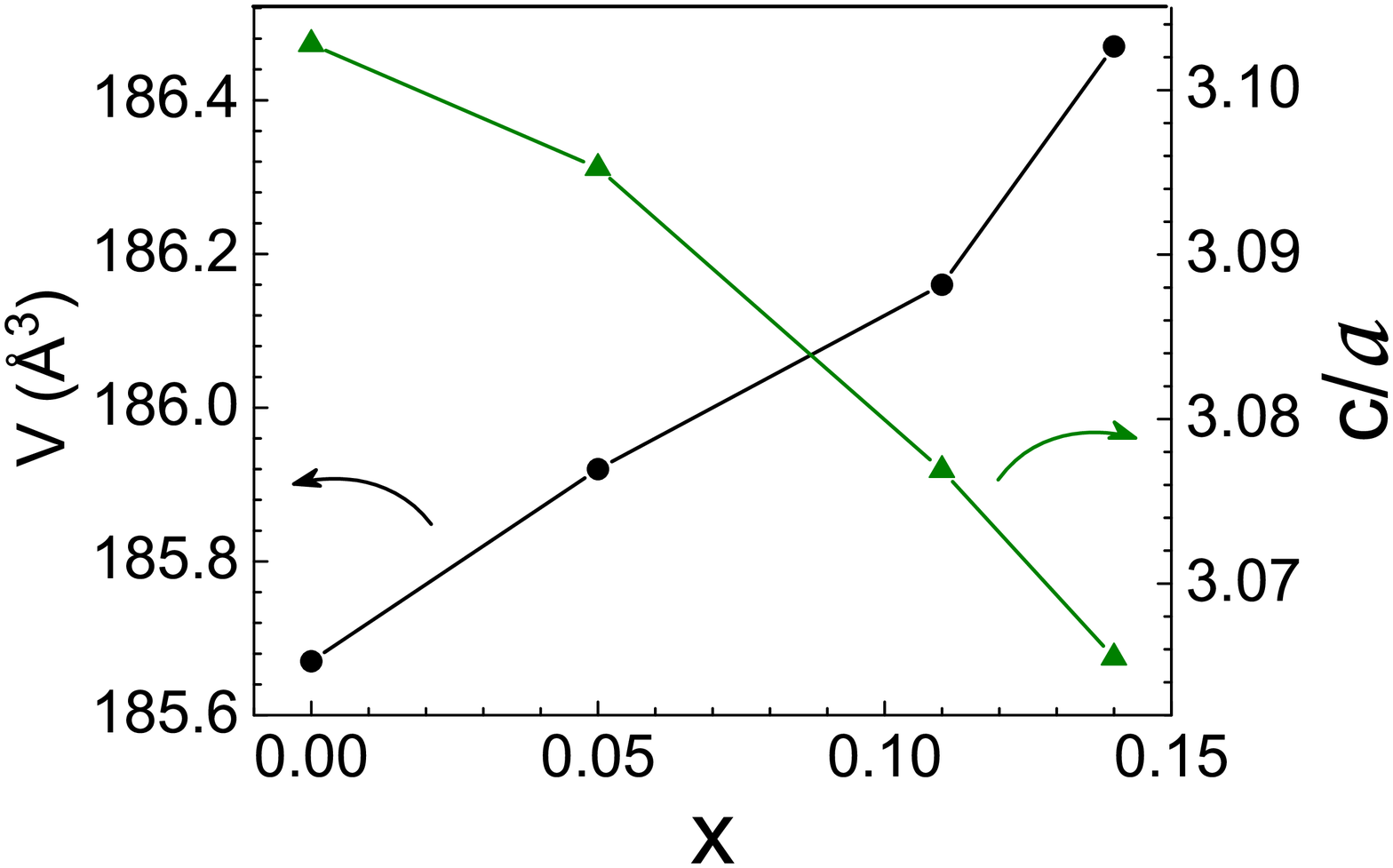}
\caption{\label{fig:XRD} (Color online) (a) The powder x-ray diffraction pattern of Eu(Fe$_{1-x}$Ir$_{x}$)$_{2}$As$_{2}$ ($x$ = 0, 0.05, 0.11 and 0.14) recorded at room temperature. (b) and (c) shows the doping dependence of lattice parameters.}
\end{figure}

The room temperature powder x-ray diffraction patterns (Fig. 1.a) for the Eu(Fe$_{1-x}$Ir$_{x}$)$_{2}$As$_{2}$ ($x$ = 0, 0.05, 0.11 and 0.14) samples reveal that all the samples crystallize in ThCr$_{2}$Si$_{2}$-type tetragonal crystal structure (space group \textit{I4/mmm}).  Single phase nature of the samples are evident along with a very small amount of FeAs impurity phase\cite{Ren1} which can be removed by optimizing the annealing process. From the EDX analysis, the atomic ratio Ir/Fe was found to be 5.45/94.55, 11.50/88.50, and 14.85/85.15 for the samples with x = 0.05, 0.11, and 0.14, respectively. The composition of the samples as revealed by EDX is very close to the starting composition. The Rietveld refinement lattice parameters are summarized in Table~I. The obtained lattice parameters of EuFe$_{2}$As$_{2}$ is in close agreement with the values as reported in literature \cite{Jeevan}. There is a continuous shift of the x-ray intensity peaks with increasing doping concentration suggesting a systematic change in the lattice parameters. Ir doping in the system results in increase of $a$-axis parameter and decrease of $c$-axis parameter leading to a reduced $c$/$a$ ratio. But the overall volume of unitcell increases with doping which is expected as Ir has higher volume than Fe. The changes of the lattice parameters and the volume of unitcell are depicted in Fig 1(b) and (c).

\begin{table}
\caption{\label{tab:XRD} Lattice parameters $a$, $c$, $c/a$ ratio and unit-cell volume $V$ of ThCr$_{2}$Si$_{2}$-type tetragonal system Eu(Fe$_{1-x}$Ir$_{x}$)$_{2}$As$_{2}$ ($x$ = 0, 0.05, 0.11 and 0.14).}
\begin{ruledtabular}
\begin{tabular}{c c c c c}
$x$ & $a$ ({\AA})  &$c$ ({\AA})		&$c/a$		& $V$ ({\AA$^3$})\\[0.5ex]
\hline
0 	 & 3.9113(2)	&12.1360(2)		&3.103		& 185.67(1)\\[1ex]

0.05 & 3.9163(1)	&12.1220(3)		&3.095		& 185.92(2)\\[1ex]

0.11 & 3.9257(3)	&12.0790(1)		&3.077		& 186.16(1)\\[1ex]

0.14 & 3.9328(3)	&12.0560(3)		&3.065		& 186.47(1)\\
\end{tabular}
\end{ruledtabular}
\end{table}

\begin{figure}[htb!]
\includegraphics[width=8.7cm, keepaspectratio]{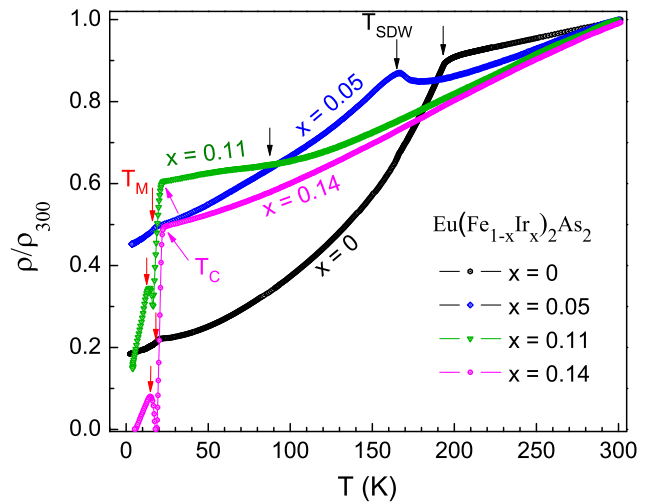}
\caption{\label{fig:Resistivity} (Color online) Temperature dependence of electrical resistivity normalized to $\rho(300 K)$ for Eu(Fe$_{1-x}$Ir$_{x}$)$_{2}$As$_{2}$ at zero field.}
\end{figure}

Fig. 2  shows the temperature dependence of electrical resistivity normalized to the value at room temperature, $\rho(T)/\rho$(300K), for the series Eu(Fe$_{1-x}$Ir$_{x}$)$_{2}$As$_{2}$ ($x$ = 0, 0.05, 0.11 and 0.14). The electrical resistivity of the parent compound EuFe$_{2}$As$_{2}$ exhibits two transitions at 190 K and 19 K corresponding to the SDW/structural transition and the antiferromagnetic ordering of Eu$^{2+}$ moments respectively\cite{Jeevan}. We find that for x = 0.05, the SDW transition is shifted towards lower temperature by 20 K and there is no signature of superconductivity. Further increasing the doping concentration to x = 0.11, a superconducting phase appears at around 21 K with a reentrant behavior in addition to SDW transition at 90 K. For a critical concentration of x = 0.14, the SDW/structural transition gets completely suppressed and a sharp drop in resistivity is observed below 22.6 K. After achieving zero value, the resistivity again starts to increase and exhibits a maximum at $\sim$ 15 K ($T_M$) and then again heads towards zero value. We attribute this to the interplay between superconductivity and the magnetic ordering of Eu$^{2+}$ moments below 18 K which hinders the superconductivity and hence the zero resistance state. This behavior is reminiscent of reentrant superconductivity observed in the ternary Chevrel phases\cite{Ishikawa} or in the rare-earth nickel borocarbides.\cite{Eisaki, Goldman, Sinha} For x $\sim$ 0.16, we find that the $T_{c}$ is shifted to 19 K and the resistivity does not reach zero value similar to that of 11\% doped sample.

\begin{figure}[htb!]
\includegraphics[width=8.8cm, keepaspectratio]{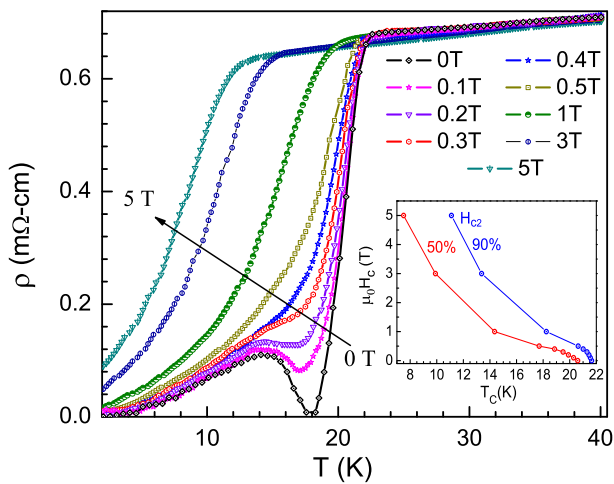}
\caption{\label{fig:RT-M} (color online) Temperature dependence of electrical resistivity of Eu(Fe$_{0.86}$Ir$_{0.14}$)$_{2}$As$_{2}$ at various applied magnetic fields. Inset: Critical fields extracted from the resistivity data at 90\% and 50\% of the normal state resistivity.}
\end{figure}

To elucidate the resistivity anomalies observed in the near optimally doped Eu(Fe$_{0.86}$Ir$_{0.14}$)$_{2}$As$_{2}$ sample, we have investigated the magnetic field dependence of resistivity (Fig. 3). With increasing applied magnetic field, the resistivity drop shifts towards the lower temperature and becomes broadened, confirming the SC transition. On the other hand, the reentrant feature of the electrical resistivity gets smeared out with increasing magnetic field. For H $\geq$ 1 T, only a broadened superconducting transition is seen competing with the ferromagnetic component of the Eu$^{2+}$ moment ordering which prevents the resistivity to attain zero value. The upper critical fields H$_{c}(T)$ vs $T_{c}$ is shown in the inset of Fig. 3 where $T_{c}$ is defined at resistivity values corresponding to 90\% and 50\% of the normal state resistivity. The behavior of H$_{c}(T)$ is as expected for superconductors in presence of magnetic ions showing a magnetic phase transition below $T_{c}$. Here, the Eu$^{2+}$ magnetic transition around 18 K is clearly influencing in change of slope in H$_{c}(T)$. Similar scenario has been observed in case of Co doped EuFe$_{2}$As$_{2}$ system\cite{Jiang} whereas a deep minimum of H$_{c}(T)$ appears around the antiferromagnetic transition temperature of rare-earth ions in RNi$_{2}$B$_{2}$C (R = rare-earth).\cite{Schmidt} Here we have made a rough estimation of upper critical field H$_{c2}$(0) using the Werthamer-Helfand-Hohenberg formula  H$_{c2}$(0)=-0.693\,$T_{c}$($\partial$H$_{c2}$/$\partial$$T$)$\mid_{T=T_c}$ where we have used the initial slope  $\mu_{0}$($\partial$H$_{c2}$/$\partial$$T$)= -0.95 T/K, yielding an upper critical field of $\sim$ 15 T. This upper critical filed is lower than the Pauli Paramagnetic limit  $\mu_{0}$H$_{P}$=1.84$T_{c}$=39.7 T. The lower value of H$_{c2}$(0) in Eu-containing superconductors, for instance in Co doped EuFe$_{2}$As$_{2}$ (H$_{c2}$(0) = 26 T)\cite{Jiang} and P doped EuFe$_{2}$As$_{2}$ (H$_{c2}$(0) = 30 T)\cite{Qian}, suggests a presence of significant internal magnetic field from Eu$^{2+}$ moments.

\begin{figure}[htb!]
\includegraphics[width=8.8cm, keepaspectratio]{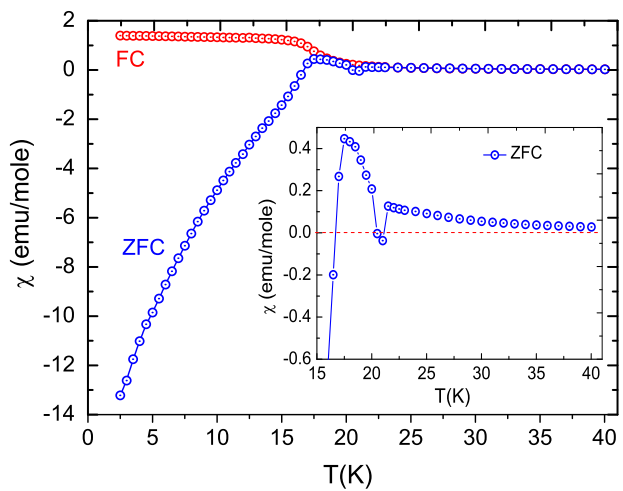}
\caption{\label{fig:susceptibility} (Color online) Temperature dependence of ZFC and FC dc magnetic susceptibility of Eu(Fe$_{0.86}$Ir$_{0.14}$)$_{2}$As$_{2}$ under applied magnetic field of 5 Oe. Inset: enlarged view of the ZFC data around $T_{C}$. }
\end{figure}
Further evidence of the reentrant superconductivity is obtained from the temperature dependence of dc magnetic susceptibility on the near optimally doped sample Eu(Fe$_{0.86}$Ir$_{0.14}$)$_{2}$As$_{2}$ under an applied field of 5 Oe (Fig. 4). Due to the proximity of superconducting transition and Eu$^{2+}$ magnetic ordering, the large moment magnetic ordering of Eu$^{2+}$ dominate the superconducting diamagnetic signal which is very hard to observe directly as has been evidenced in Co doped EuFe$_{2}$As$_{2}$ system\cite{Jiang, Hu, Guguchia} or P doped EuFe$_{2}$As$_{2}$ polycrystalline samples,\cite{Qian} whereas P doped EuFe$_{2}$As$_{2}$ single crystals manifest a clear diamagnetic signal when the applied magnetic field is parallel to the ab-plane of the crystals.\cite{Jeevan1} But interestingly in our case of Ir doped EuFe$_{2}$As$_{2}$ polycrystalline system, considering ZFC data, a prominent diamagnetic signal has been observed below 16 K above which the susceptibility becomes positive and shows a maximum around 18 K. A drop in the ZFC susceptibility data at around 21 K is close to the onset $T_c$  as seen through a sharp drop in resistivity. In short the entire re-entrant behavior as observed through resistivity measurement is re-established by the low field dc magnetic susceptibility measurements.

\begin{figure}[htb!]
\includegraphics[width=8.8cm, keepaspectratio]{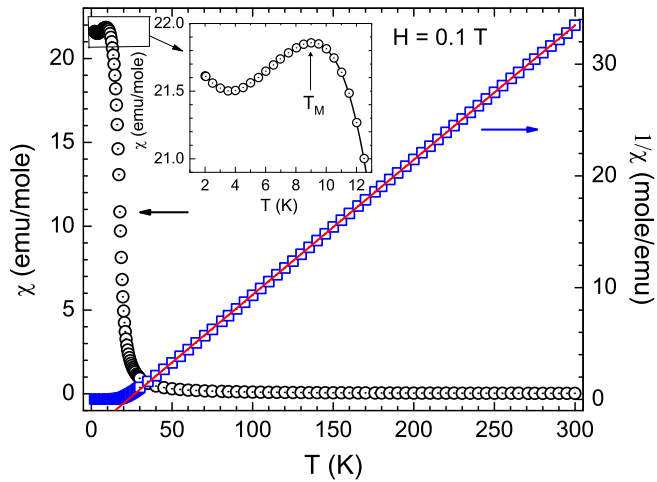}
\caption{\label{fig:susceptibility-1000}(Color online) Temperature dependence of Field cooled dc magnetic susceptibility and inverse susceptibility of Eu(Fe$_{0.86}$Ir$_{0.14}$)$_{2}$As$_{2}$ under applied magnetic field of 0.1 T. The solid line through the data points shows the fit to the modified Curie-Weiss law. Inset: enlarged view of the low temperature susceptibility.}
\end{figure}
 Fig. 5 shows the FC dc magnetic susceptibility for Eu(Fe$_{0.86}$Ir$_{0.14}$)$_{2}$As$_{2}$ under H = 0.1 T. The $\chi(T)$ data could be well described by the modified Curie-Weiss law $\chi(T)= \chi_{0}+ C/(T - \theta)$ above 30 K, where $\chi_{0}$ represents the temperature independent term, $C$ is the Curie constant and $\theta$ is the Curie-Weiss temperature. The solid line through the data points in Fig. 5 shows the fit to this law with the fitting parameters $\chi_{0}= 1.86\times 10^{-4}$~emu/mole, $C$ = 8.23 emu K/mole and $\theta$ = 22.8 K. The calculated value of effective magnetic moment $\mu_{eff}$ = 8.1 $\mu_{B}$ per Eu-ion which is close to the theoretical value of 7.94 $\mu_{B}$ for free Eu$^{2+}$ ion with $S = 7/2$. Below 20 K, $\chi(T)$ increases steeply with decreasing temperature, however, it does not saturate at lower temperature as shown in the inset of Fig. 5. Furthermore, the temperature dependence of zero filed cooled (ZFC) and field cooled (FC) dc magnetization measurements have been performed for Eu(Fe$_{0.86}$Ir$_{0.14}$)$_{2}$As$_{2}$ under various fixed magnetic fields. The data reveal a decrease of $T_{M}$ (the temperature at which the magnetization shows a maximum) with increasing applied magnetic field up to 0.1 T [Fig. 6]. For H $\geq$ 0.2 T, the low temperature magnetization appears to be more of a field stabilized FM phase. Under high magnetic field, the saturation of magnetization gives a fully polarized value $\sim$ 6.9 $\mu_{B}$/f.u. as expected for parallelly aligned Eu$^{2+}$ moments (g$S$ = 7.0 $\mu_{B}$/f.u. with g = 2, $S = 7/2$). The FC and ZFC magnetization data coincide at higher temperatures, they differ significantly only at lower temperature below $T_{M}$ where ZFC magnetization values are lower than those of FC values. All the above observations suggest that the magnetic ordering could be antiferromagnetic with the presence of a ferromagnetic component, so called canted antiferromagnet (C-AFM). Co doped EuFe$_{2}$As$_{2}$ system\cite{Guguchia} shows similar magnetic behavior where Eu$^{2+}$ moments are found to be ordered C-AFM which causes reentrance in the superconductivity. On the other hand, in P doped EuFe$_{2}$As$_{2}$ system,\cite{Jeevan1} Eu$^{2+}$ moments order antiferromagntically which coexists with superconductivity but above a certain doping level Eu$^{2+}$  moments order ferromagnetically which destroys the superconductivity. Later, the same group came up with a detailed discussion proposing that the magnetic moments of the Eu$^{2+}$ ions are not simply AFM aligned in adjacent ab planes, but canted, yielding a ferromagnetic contribution along the c direction and the superconducting phase coexists with the C-AFM in EuFe$_{2}$(As$_{1-x}$P$_{x}$)$_2$ crystals.\cite{Zapf}

\begin{figure}[htb!]
\includegraphics[width=8cm, keepaspectratio]{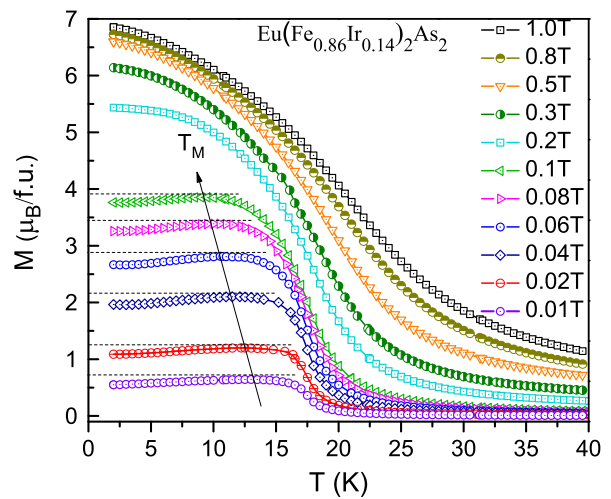}
\caption{\label{fig:susceptibility} (Color online) Temperature dependence of ZFC magnetization of Eu(Fe$_{0.86}$Ir$_{0.14}$)$_{2}$As$_{2}$ measured at various fixed magnetic fields. Dotted horizontal lines are drawn to show the decrease of M below the maxima.}
\end{figure}

To get some information about the electronic states for the doped compound, we carried out the density-functional band structure calculations using full potential linear augmented plane wave plus local orbitals (FP-LAPW+lo) method as implemented in WIEN2K code.\cite{Blaha} Pedrew-Burke-Ernzerhof (PBE) form of the generalized gradient approximation (GGA) was employed for the exchange correlation potential.\cite{Perdew} Additionally, to account for the strong Coulomb repulsion within the Eu $4f$ orbitals we have included U on a mean-field level using the GGA+U approximation. There exists no spectroscopy data for EuFe$_{2}$As$_{2}$, therefore, we have used U = 8 eV, the standard value for an Eu$^{2+}$ ion.\cite{Jeevan, Hossain1, Li} We have employed supercell calculations to explicitly study the effect of partial Ir substitutions on Fe site. For this purpose, $2\times2\times1$ supercell of the EuFe$_2$As$_2$ unit cell was constructed for a 12.5\% Ir doped (in between 11\% and 14\% doping concentration for which superconductivity evolved) compound. We then replaced one Fe in each plane by Ir which corresponds to the nominal composition of Eu$_{8}$Fe$_{14}$Ir$_{2}$As$_{16}$. The lattice constants for the unitcell of 12.5\% Ir doped compound was obtained by interpolating linearly the experimental lattice parameters listed in Table 1, outcome of which is $a$ = 3.930 $\AA$ and $c$ = 12.066 $\AA$. The atomic position of As was kept fixed at $z$ = 0.362 (experimental $z_{As}$ of EuFe$_{2}$As$_{2}$) for both the compounds.

\begin{figure}[htb!]
\includegraphics[width=8.5cm, keepaspectratio]{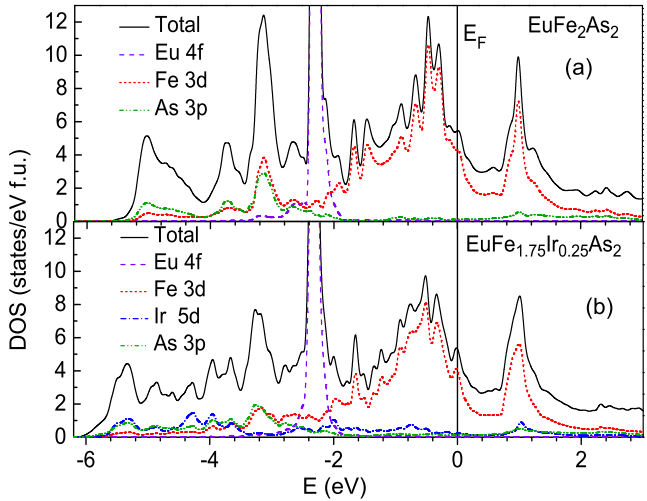}
\caption{\label{fig:DOS} (color online) The total DOS and partial DOS per f.u. of (a) EuFe$_{2}$As$_{2}$ and (b) EuFe$_{1.75}$Ir$_{0.25}$As$_{2}$ in the non- magnetic state in Fe layer and FM interaction between the intralayer Eu spins in Eu layer. The unfilled Eu $4f$ states are at about 10 eV above the Fermi level. The Fermi level (E$_{F}$) corresponds to zero binding energy.}
\end{figure}

We have performed electronic structure calculations for both EuFe$_{2}$As$_{2}$ and EuFe$_{1.75}$Ir$_{0.25}$As$_{2}$ in the quenched paramagnetic state, that means no spin polarization is allowed on the Fe or Ir ions in the calculations. The general shape of our density of states for EuFe$_{2}$As$_{2}$ (Fig. 7.a) is similar to that reported in literature.\cite{Jeevan, Li} Since the Eu $4f$ states for both the compounds are quite localized, the Eu ions are in a stable 2+ valence state. The calculated spin moment for Eu$^{2+}$ is about 6.9 $\mu_{B}$ for both the compounds which is consistent with the experimental values. Apart from the Eu $4f$ states, the remaining density of states (DOS) is modified significantly due to the Ir substitution. More importantly, near the Fermi level, the DOS for EuFe$_{2}$As$_{2}$ is almost flat as observed for other iron-pnictide parent compounds\cite{Nekrasov} whereas for Ir doped compound, a peak appears near the Fermi level. Similar effect has been observed for a Ir doped SrFe$_{2}$As$_{2}$ superconducting system\cite{Lijun} or K doped EuFe$_{2}$As$_{2}$ system.\cite{Hossain1} Total DOS near the Fermi level is mainly contributed by Fe $3d$ states but a small contribution comes from the Ir $5d$ character as well. The total DOS at the Fermi level N[E$_{F}$] is 5.48 states/eV f.u for EuFe$_{2}$As$_{2}$ which is reduced to 5.24 states/eV f.u for EuFe$_{1.75}$Ir$_{0.25}$As$_{2}$ . Our calculation on the doped compound shows a shift in Fermi energy by 0.02 eV above the Fermi level and a change in band filling as compared to the pure compound, indicating electron doping in the system by partial substitution of Fe by Ir. As can be seen from Fig. 7 , the overall DOS for the doped compound is reduced throughout the energy scale as compared to pure EuFe$_{2}$As$_{2}$. The reduction in DOS is associated with the extended d-band width and stronger hybridization involving Ir. The occurrence of superconductivity in the doped system can also be justified by the change in structural parameters and increased hybridization due to Ir substitution. Since the $c$ lattice parameters shrink rather anisotropically i.e. $c/a$ decreases with increasing Ir content (see Table I), the lattice becomes more three-dimensional. Similar anisotropic change in the lattice parameters leading to bulk SC were also observed for isovalent substitution of Fe by larger Ru atoms or As by smaller P atoms in AFe$_{2}$As$_{2}$ (A = Sr, Eu).\cite{Schnelle, Jeevan1} In these systems, the substitution does not provide electron or hole doping but rather suppresses the Fe SDW order in favor of superconductivity by reducing the Stoner enhancement of Fe and increasing bandwidth due to stronger hybridization. So, apart from the electron doping, Ir substitution has similar effect to that of pressure i.e. broadening the bands and increasing hybridization which also plays an important role in suppressing SDW and inducing superconductivity in the system.

In summary, reentrant superconductivity has been evidenced in Eu(Fe$_{1-x}$Ir$_{x}$)$_{2}$As$_{2}$ samples through our comprehensive investigation of (magneto)resistivity and low-field magnetic susceptibility. With increasing Ir doping,  the SDW order in EuFe$_{2}$As$_{2}$ is gradually suppressed and superconductivity is induced at $\approx$ 22.6 K for 14\% Ir doped sample. The low field magnetization measurements of Eu(Fe$_{0.86}$Ir$_{0.14}$)$_{2}$As$_{2}$ show a prominent diamagnetic signal due to superconductivity with a reentrant feature. The magnetization measurements at various applied magnetic fields reveal that the the magnetic ordering temperature ($T_{M}$) of Eu$^{2+}$ moments  shifts towards lower temperature with increasing field up to 0.1 T and above which the magnetization looks like more of a field stabilized FM phase. Thus, below 18 K, Eu$^{2+}$ moments are most likely ordered antiferromagnetically with the presence of a ferromagnetic component (canted antiferromagnet) which causes resistivity reentrance. Further experiments are planned to probe the magnetism of Eu ions as well as to investigate the possible presence of spontaneous vortices.

This work has been supported by the Council of Scientific and Industrial Research, New Delhi (Grant No. 80(0080)/12/ EMR-II).

\end{document}